\documentclass[twocolumn]{aastex6}
\usepackage{amsmath}

\fullcollaborationName{}

\def\simgt{\lower.5ex\hbox{\gtsima}} 
\def\simlt{\lower.5ex\hbox{\ltsima}} 
\def\gtsima{$\; \buildrel > \over \sim \;$} 
\def\ltsima{$\; \buildrel < \over \sim \;$}
\def\Msun{M_\odot}

\newcommand\lsim{\mathrel{\rlap{\lower4pt\hbox{\hskip1pt$\sim$}}
        \raise1pt\hbox{$<$}}}
\newcommand\gsim{\mathrel{\rlap{\lower4pt\hbox{\hskip1pt$\sim$}}
        \raise1pt\hbox{$>$}}}
\def\myputfigure#1#2#3#4#5%
{\vskip#5pt\makebox[0pt]{\hskip#2in
\includegraphics[width=#3\textwidth]{#1}}\vskip#4pt\hfill}

\begin{document}

\title[Conditions for Optimal Growth of Black Hole Seeds]
      {Conditions for Optimal Growth of Black Hole Seeds}
\author{Fabio Pacucci\altaffilmark{1,2}, Priyamvada Natarajan\altaffilmark{1,2,3}, Marta Volonteri\altaffilmark{4}, Nico Cappelluti\altaffilmark{1,2}, C. Megan Urry\altaffilmark{1,2,3}}
\and

\altaffiltext{1}{Department of Physics, Yale University, New Haven, CT 06511, USA.}
\altaffiltext{2}{Yale  Center  for  Astronomy  and  Astrophysics, Yale University, New Haven, CT 06520, USA.} 
\altaffiltext{3}{Department of Astronomy, Yale University, New Haven, CT 06511, USA.}  
\altaffiltext{4}{Institut d'Astrophysique de Paris, Sorbonne Universit\'es, UPMC Univ. Paris 06 et CNRS, UMR 7095, F-75014, Paris, France.}

\label{firstpage}
             
\begin{abstract}
Super-massive black holes weighing up to $\sim 10^9 \, \mathrm{\Msun}$ are in place by $z \sim 7$, when the age of the Universe is $\lesssim 1 \, \mathrm{Gyr}$. This implies a time crunch for their growth, since such high masses cannot be easily reached in standard accretion scenarios. 
Here, we explore the physical conditions that would lead to optimal growth wherein stable super-Eddington accretion would be permitted. Our analysis suggests that the preponderance of optimal conditions depends on two key parameters: the black hole mass and the host galaxy central gas density. In the high-efficiency region of this parameter space, a continuous stream of gas can accrete onto the black hole from large to small spatial scales, assuming a global isothermal profile for the host galaxy. Using analytical initial mass functions for black hole seeds, we find an enhanced probability of high-efficiency growth for seeds with initial masses $\gsim 10^4 \, \mathrm{\Msun}$.  
Our picture suggests that a large population of high-$z$ lower-mass black holes that formed in the low-efficiency region, with low duty cycles and accretion rates, might remain undetectable as quasars, since we predict their bolometric luminosities to be $\lesssim 10^{41} \, \mathrm{erg \, s^{-1}}$. The presence of these sources might be revealed only via gravitational wave detections of their mergers.
\end{abstract}


\setcounter{footnote}{1}
\newcounter{dummy}

%%%%%%%%%%%%%%%%%%%%%%%%%%%%%%%%%%%%%%%%%%%%%%%%%%%%%%%%%%%%%%
%% SECTION 1: INTRODUCTION
%%%%%%%%%%%%%%%%%%%%%%%%%%%%%%%%%%%%%%%%%%%%%%%%%%%%%%%%%%%%%%
\section{Introduction}
\label{sec:introduction}
The observation of a few super-massive black holes (SMBHs) by redshift $z \sim 7$ \citep{Fan_2006,Mortlock_2011, Wu_2015} has deeply challenged our view of black hole formation and growth. These collapsed objects with masses up to $\sim 10^{9-10} \, \mathrm{\Msun}$ appear to be in place less than $1 \, \mathrm{Gyr}$ after the Big Bang.
These observations opened up a seeding problem, where the word ``seed" refers to the original black hole that, by accretion and merging, generates a SMBH.
These final masses, in fact, are hard to achieve with seeds originating from the end states of stars.
In the standard scenario, the black hole mass $M_{\bullet}$ increases exponentially, with a time scale $t_{S} \sim 0.045\epsilon_{0.1} \, \mathrm{Gyr}$, with $\epsilon_{0.1}$ the matter-radiation conversion factor normalized to the standard value of $10\%$:
\begin{equation}
M_{\bullet}(t) = M_{\bullet ,0} \exp\left({\frac{t}{t_S}}\right) \, .
\label{eq:growth}
\end{equation}
The problem is manifest if we start the growth from a standard seed with an initial mass $M_{\bullet} \lesssim 100 \, \mathrm{\Msun}$, formed at the endpoint of the evolution of a massive star.
This remnant mass is relevant both for the formation in the local Universe (Pop I/II stars) as well as for the formation in the early Universe, with more massive Pop III stars.
In the context of a light seeding model from stellar remnant seeds, this growth process requires a \textit{constant} accretion at the Eddington rate until $z \sim 7$ to allow the buildup of $\sim 10^9 \, \mathrm{\Msun}$ SMBHs.
Though not impossible, this process is unlikely, due to the stringent requirement of a steady gas reservoir to maintain this supply rate.
To alleviate the problem, a twofold solution can be mathematically devised from Eq.~\ref{eq:growth}: either decreasing the time-scale $t_S$ or increasing the initial mass $M_{\bullet ,0}$.

The first possibility is achieved by decreasing the radiative efficiency $\epsilon$, or in other words, by assuming that accretion rates are not bounded by the Eddington limit \citep{Volonteri_2005}.
Several works (\citealt{Alexander_2014, Madau_2014, Volonteri_2014}) have predicted the occurrence of largely super-Eddington accretion episodes at high redshift, with rates several times larger than the Eddington accretion rate $\dot{M}_{Edd}=L_{Edd}/c^2$.
Mass supply rates available in high-$z$ galaxies allow for episodes of super-Eddington accretion. Assuming a velocity dispersion $\sigma$ ($\sigma_{100}$ is expressed in units of $100 \, \mathrm{km \, s^{-1}}$) for the dark matter halo hosting a black hole, the characteristic freefall rate of self-gravitating gas exceeds the Eddington rate by a factor \citep{Begelman_2017}:
\begin{equation}
\dot{M}_{ff} \sim 10^5 \sigma_{100}^3 \dot{M}_{\rm Edd} \left( \frac{M_{\bullet	}}{10^6 \, \mathrm{\Msun}}\right)^{-1} \, .
\end{equation}
The problem is not the availability of gas, but the efficiency in streaming the gas from the large-scale reservoir down to sufficiently small scales to be promptly accreted.

The second possibility assumes that the environmental conditions of the early Universe ($z \gsim 10-15$) allowed alternative pathways to form massive black hole seeds: (i) the direct collapse of unenriched and self-gravitating pre-galactic disks (\citealt{Begelman_2006, Lodato_Natarajan_2006, Lodato_2007}), (ii) the collapse of a primordial atomic-cooling halo into a direct-collapse black hole (DCBH, \citealt{Bromm_Loeb_2003, Shang_2010, Johnson_2012, Natarajan_2017}), or (iii) the formation of a very massive star from runaway stellar mergers in a dense cluster (\citealt{Devecchi_2009, Davies_2011, Stone_2017}). 

In this Letter we investigate the physical conditions that optimize the growth of black hole seeds, \textit{independent of the seeding mechanism}. The only relevant parameter is the mass of the black hole, not the mechanism that led to its formation. We explore the conditions for optimal growth, defined as an accretion process that minimizes the time needed to reach a given black hole mass $M_{{\bullet}}$, starting from a seed mass $M_{{\bullet},0}$.
We provide a global view of the physical conditions that permit a stable super-Eddington accretion, from small to large spatial scales, including a careful treatment of the angular momentum of the gas.

%%%%%%%%%%%%%%%%%%%%%%%%%%%%%%%%%%%%%%%%%%%%%%%%%%%%%%%%%%%%%%%%
%% SECTION 2: 
%%%%%%%%%%%%%%%%%%%%%%%%%%%%%%%%%%%%%%%%%%%%%%%%%%%%%%%%%%%%%%%%
\section{Efficiency of Black Hole Growth \\ on Different spatial scales}
\label{sec:previous}
We start by describing the physical effects that influence optimal conditions for black hole growth. \textit{Optimal growth conditions are reached whenever the combined effect of radiation pressure and angular momentum is ineffective in halting the accretion flow, both at small and at large spatial scales}. This requires that the flow is unimpeded from the largest to the smallest scales, which occurs for a set of conditions in the host halo.
Our analysis of the coupling between the black hole seed and its host halo is investigated in a two-dimensional parameter space.
The parameters are the seed mass, $M_{\bullet}$, and for the host halo, the number density of the gas, $n_{\infty}$, at the Bondi radius, $R_B = 2 G M_{\bullet}/c_s^2$, \citep{Bondi_1952}, where $c_s$ is the sound speed of the unperturbed gas surrounding the black hole.
%In addition, we impose the conditions required for the flow down to the smallest scales modulo the assumption that the density profile of the gas in the halo can be approximated via an isothermal.

In the following we describe how three conditions for optimal growth of the black hole seed can be expressed in this parameter space, given reasonable assumptions for the angular momentum of the halo.
We assume an isothermal density profile for the gas in the host galaxy:
\begin{equation}
\rho(r) = \frac{f_g \sigma^2}{2 \pi G r^2} \, ,
\label{eq:density}
\end{equation}
where $f_g = \Omega_b/\Omega_m = 0.16$ is the cosmological gas fraction and the velocity dispersion $\sigma$ is linked to the halo mass, following \cite{Pacucci_2017_maxmass}.

\subsection{Growth Efficiency on Small Scales}

The growth efficiency on small scales ($r \ll R_B$) is determined by the extent of the transition radius $R_T$ \citep{PVF_2015}, the spatial scale above which the accretion flow is dominated by radiation pressure that powers outflows.
The definition of $R_T$ follows from the comparison of two relevant time scales, the feedback ($t_{\mathrm{fb}}$) and the accretion ($t_{\mathrm{acc}}$) times. The former, computed at some radius $r$, is the e-folding time needed by the radiation pressure to alter $\dot{M}$: 
\begin{equation}
\dot{M}_{\bullet}(t) = \dot{M}_{\bullet}(t=0) e^{-t/t_{\mathrm{fb}}} \, .
\end{equation}
The latter is the time needed to accrete the gas mass $M_g$ inside a sphere of radius $r$:
\begin{equation}
t_{\mathrm{acc}} = \frac{M_g(<r)}{\dot{M}_{\bullet}} \, .
\end{equation}
At the transition radius: $t_{\mathrm{fb}}(R_T) = t_{\mathrm{acc}}(R_T)$.

$R_T$ then defines and demarcates the two radial accretion regimes as noted previously by \cite{Park_2016}: a feedback-limited growth regime ($r \gg R_T$), wherein radiative feedback is important, and a gas-supply-limited one ($r \ll R_T$), when radiative feedback is unimportant and most of the available gas is readily accreted.
The transition radius needs to be compared with the typical spatial scale where additional forces, other than gravity and radiation pressure, come into play and favors the accretion of gas (e.g., the photon-trapping radius; see \citealt{Begelman_1979}). 
This comparison provides the crucial additional condition for efficient growth at small scales in the $M_{\bullet}-n_{\infty}$ parameter space.
Comparing the transition radius to the trapping radius, this small-scale constraint translates into the following condition \citep{PVF_2015}:
\begin{equation}
M_{\bullet} > 10^{-11}  \left(\frac{n_{\infty}}{1 \, \mathrm{cm^{-3}}}\right)^2 \, \mathrm{\Msun} \, .
\label{eq:small_scales_final}
\end{equation}
Eq. \ref{eq:small_scales_final} suggests that an increase in the gas number density in the host galaxy, $n_{\infty}$, leads to an increase in the minimum seed mass necessary to sustain efficient growth. An increase in $n_{\infty}$ leads to the possibility of building up a larger accretion rate: $\dot{M} \propto n_{\infty}$, by definition. If the black hole is more massive, then its Eddington accretion rate is larger. In this event the central object could sustain larger absolute accretion rates without reaching the Eddington limit.

\subsection{Growth Efficiency on Large Scales}

\cite{Inayoshi_2016_super} investigated the conditions that lead to efficient growth on large spatial scales, $r \gsim R_B$.
They studied spherically symmetric accretion flows onto massive black holes (in the mass range $10^2 \, \mathrm{M_{\odot}} < M_{\bullet} <10^6  \, \mathrm{M_{\odot}}$) embedded in dense metal-poor clouds.
They found that rapid gas supply from the Bondi radius can reach super-Eddington rates above a certain density:
\begin{equation}
\left( \frac{n}{10^5 \mathrm{cm^{-3}}}\right) > \left(\frac{M_{\bullet}}{10^4 \, \mathrm{\Msun}}\right)^{-1} \left(\frac{T}{10^4 \, \mathrm{K}}\right)^{3/2} \, .
\label{eq:large_scales}
\end{equation}

The solution comprises a core, where photon trapping is relevant, and an accreting external region that follows a Bondi profile. Since the effect of photon trapping is important, the radiation from the core does not significantly modify the gas dynamics at larger scales and super-Eddington accretion rates can be sustained.
Assuming gas at the atomic-cooling threshold $T \sim 10^4 \, \mathrm{K}$, Eq.~\ref{eq:large_scales} can be turned into a relation between $M_{\bullet}$ and $n_{\infty}$:
\begin{equation}
M_{\bullet} > 10^9 \left( \frac{n_{\infty}}{\mathrm{1 \, cm^{-3}}} \right)^{-1} \, \mathrm{\Msun} \, .
\label{eq:large_scales_final}
\end{equation}
An increase in $n_{\infty}$ leads to a decrease in the minimum seed mass necessary to sustain efficient growth. In fact, larger column densities of gas may lead to larger accretion rates, which in turn make the effects of photon trapping more relevant \citep{Begelman_1979}. The H II region around the black hole thus remains smaller than the Bondi radius, leading to easier acceptance of large quantities of gas from large to smaller spatial scales.
We note that the assumption that the gas is at the atomic-cooling threshold is valid only in the high-$z$ Universe.

\subsection{Growth Efficiency and Angular Momentum}

The angular momentum barrier needs to be overcome by the infalling gas before it can accrete onto the black hole.
\cite{Begelman_2017} investigated the conditions that allow efficient accretion with specific angular momentum of gas at the Bondi radius, $\ell_{\rm B}$.
The parameter that controls the accretion efficiency is the ratio of the specific angular momentum of the gas to its Keplerian value computed at the trapping radius: $\lambda_{\rm B} = \ell_{\rm B}/(GMR_{\rm B})^{1/2}$.
The trapping radius, $R_{\rm tr}$ \citep{Begelman_1979}, is the distance from the accreting black hole inside which photon trapping is efficient: $R_{\rm tr} = (\dot{M}/\dot{M}_{\rm Edd}) R_g$, where $R_g = GM/c^2$ is the gravitational radius of the black hole.
The condition for optimal growth is:
\begin{equation}
\lambda_{\rm B} < 0.1 \sigma_{100}^{5/2} \left( \frac{\dot{M}}{\dot{M}_{ff}} \right)^{1/2} \left( \frac{M_{\bullet}}{\mathrm{10^6 \, \Msun}} \right)^{-1/2} \, .
\end{equation}
The critical value of $\lambda_{\rm B}$ is defined as $\lambda_{\rm B, crit}$.
This condition follows by requiring that the centrifugal barrier fall inside the trapping radius. Angular momentum is thus deposited under highly trapped conditions and the radiative pressure is ineffective in pushing the gas away.

Expressing the previous condition in terms of $M_{\bullet}$ and $n_{\infty}$, with $\lambda_{\rm B}$ as a free parameter, we obtain:
\begin{equation}
M_{\bullet} > 2.2 \times 10^{19} \left( \frac{n_{\infty}}{\mathrm{1 \, cm^{-3}}} \right)^{-5/4} \lambda_{\rm B}^{3} \, \mathrm{\Msun} \, .
\label{eq:angmom_final}
\end{equation}

\subsection{Conditions for the Optimal Growth of Seeds}
\begin{figure*}
\centering
\includegraphics[angle=0,width=0.45\textwidth]{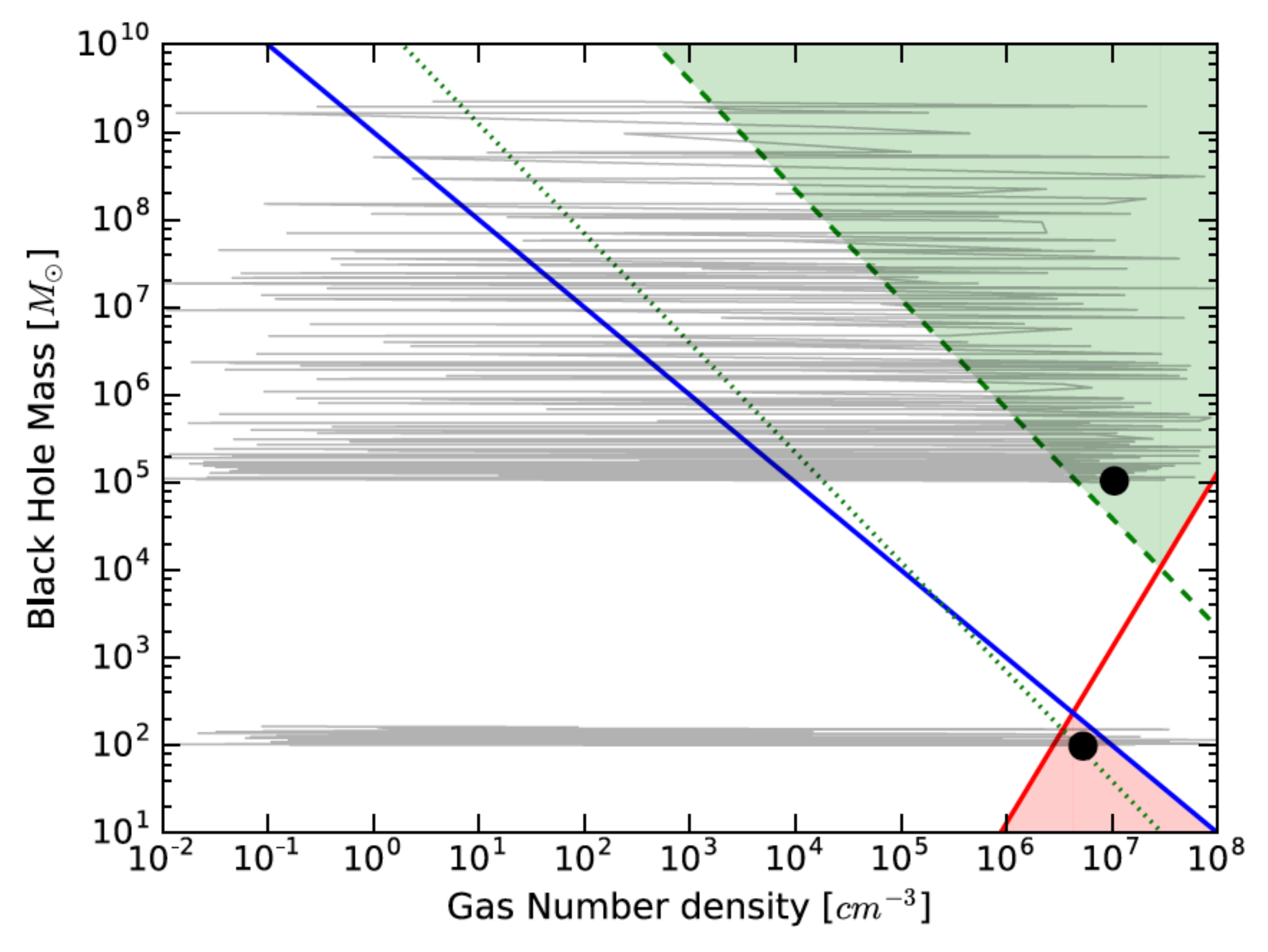}
\includegraphics[angle=0,width=0.45\textwidth]{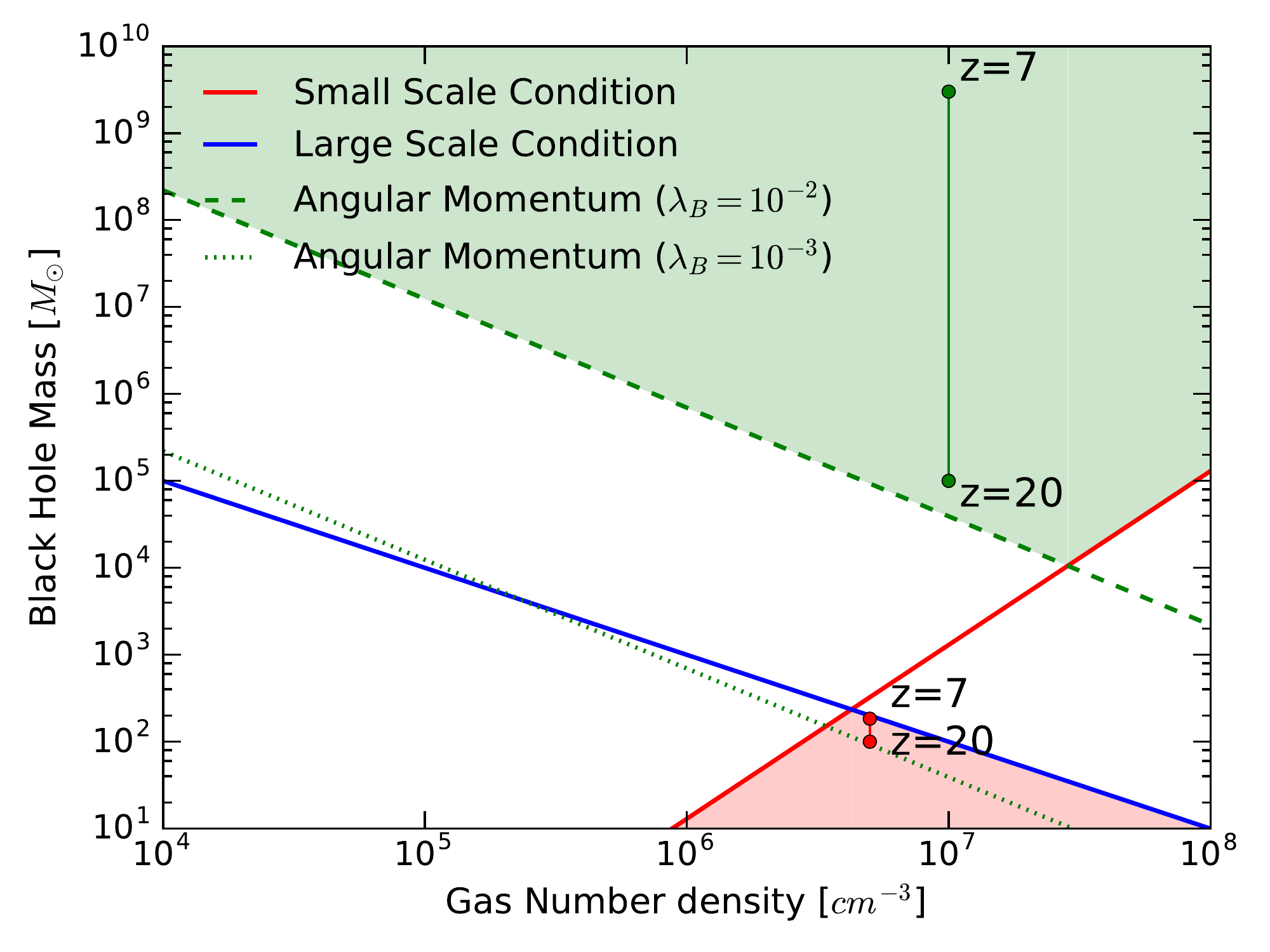}
\vspace{-0.1cm}
\includegraphics[angle=0,width=0.5\textwidth]{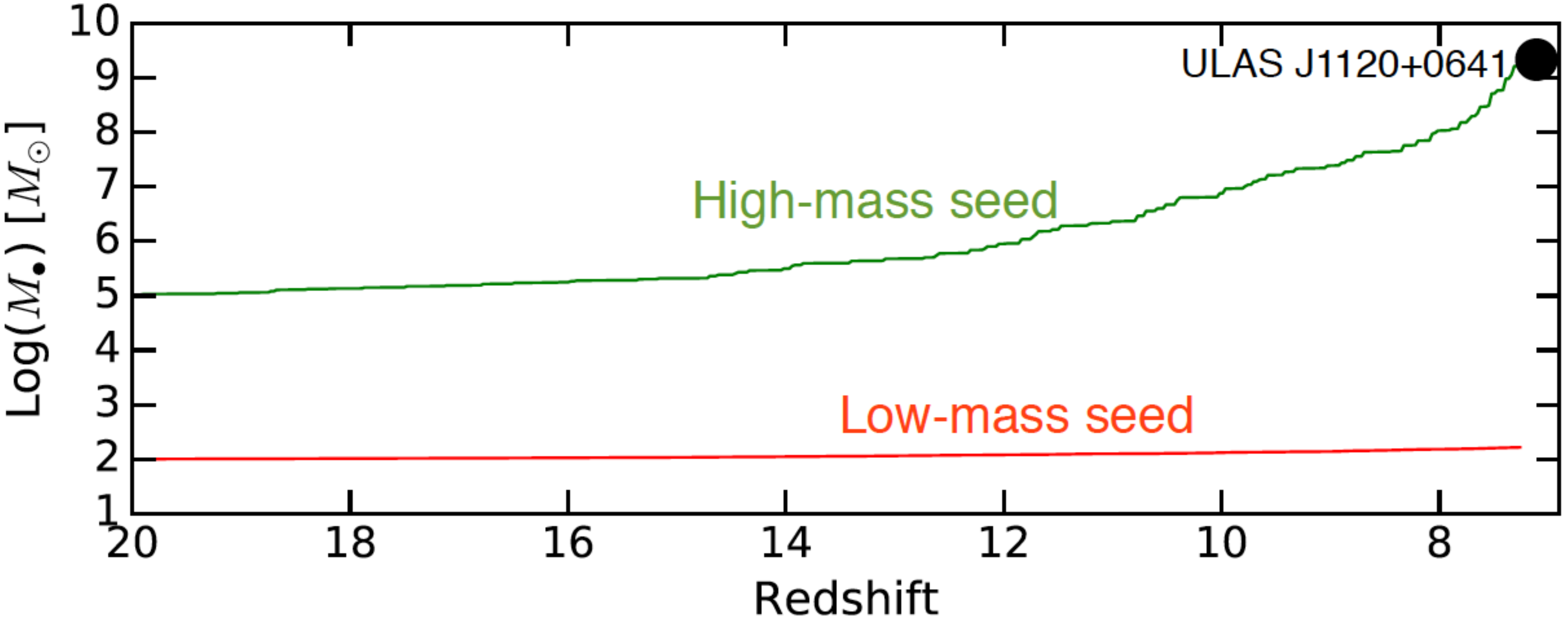}
\caption{Conditions for optimal black hole growth in the $M_{\bullet}-n_{\infty}$ parameter space. The green shaded area is the high-efficiency region, in which all the three conditions are met and where accretion is efficient.
Accretion outside this region occurs at low efficiency. In particular, in the low-efficiency region we identify a ``super-bad" area, shaded in red, where efficient gas accretion is impeded by all the conditions.
\textbf{Left panel:} we show an example of cosmological evolution from $z_0=20$ to $z=7$ of two black hole seeds (their initial conditions are shown as black circles), with initial masses $M_{\bullet,1} = 10^2 \, \mathrm{\Msun}$ and $M_{\bullet,2} = 10^5 \, \mathrm{\Msun}$, embedded in galaxies with initial mean gas number densities $n_{\infty,1} = 5 \times 10^6 \, \mathrm{cm^{-3}}$ and $n_{\infty,2} = 10^7 \, \mathrm{cm^{-3}}$. The black hole mass is evolved with duty cycles and accretion rates in accordance with the instantaneous position in the parameter space, while the densities are evolved in a random fashion between $10^{-2} \, \mathrm{cm^{-3}}$ and $10^8 \, \mathrm{cm^{-3}}$.
The ``lucky" seed easily reaches $\sim 2 \times 10^{9} \, \mathrm{\Msun}$ by $z \sim 7$. \textbf{Right panel:} only the mass evolution of the left panel is shown, while the density is kept artificially at the initial value for visualization clarity, and the plotted number density range is reduced to $10^{4} \, \mathrm{cm^{-3}} < n_{\infty} < 10^8 \, \mathrm{cm^{-3}}$. The red (green) line indicates inefficient (efficient) growth. \textbf{Bottom panel:} we illustrate the mass buildup of the two seeds as a function of the redshift. The data point corresponds to the highest-redshift quasar ever discovered \citep{Mortlock_2011}.}
\label{fig:super-super}
\end{figure*}

The right panel in Fig.~\ref{fig:super-super} shows the relations presented above 
(Eqs. \ref{eq:small_scales_final}, \ref{eq:large_scales_final}, \ref{eq:angmom_final}) 
in the $M_{\bullet}-n_{\infty}$ parameter space. The condition on the radiation pressure at large scales becomes relevant only with low values of the angular momentum.
While Eqs.~\ref{eq:large_scales_final} and \ref{eq:angmom_final} show a decrease of the minimum mass for increasing $n_{\infty}$, the relation for small scales (Eq.~\ref{eq:small_scales_final}) has an opposite trend.
We designate the ``high-efficiency region" as the area of the $M_{\bullet}-n_{\infty}$ parameter space in which all the three conditions are met.
Black hole seeds transiting in this region are ``super-lucky" (or efficient accretors), since the combination of $M_{\bullet}$ and $n_{\infty}$ allows a continuous stream of gas from large to small scales. Accretion outside of this region of the parameter space is inefficient, hence the associated area is defined as the ``low-efficiency region". 
Accretion in the low-efficiency region is characterized by either low duty cycles or low accretion rates, or both.
In particular, we identify a ``super-bad" area, where efficient gas accretion is impeded by all the conditions.
{\it The region of the $M_{\bullet}-n_{\infty}$ parameter space in which the black hole seed forms determines the efficiency of its subsequent growth.} 
Here, we suggest that each point in this physical parameter space ($M_{\bullet},n_{\infty}$) can be associated with a probability, which yields a probable value for the growth efficiency as well (Sec. \ref{sec:probability}).

%%%%%%%%%%%%%%%%%%%%%%%%%%%%%%%%%%%%%%%%%%%%%%%%%%%%%%%%%%%%%%%%%%%%%%
%% SECTION 3:
%%%%%%%%%%%%%%%%%%%%%%%%%%%%%%%%%%%%%%%%%%%%%%%%%%%%%%%%%%%%%%%%%%%%%%
\section{The Cosmological Evolution of Seeds}
\label{sec:evolution}
Building on the conditions for optimal black hole growth, we predict the cosmological evolution of the black hole mass, $M_{\bullet}(z)$, starting from an initial mass $M_{\bullet,0}$ in a given halo of gas number density $n_{\infty,0}$ at high-$z$ when the parameter space explored is viable. Therefore, we can examine the early epochs of black hole growth in the context of a gas-rich Universe.
We describe the time evolution of $M_{\bullet}$ with two parameters: (i) ${\cal D}$ is the duty cycle, i.e., the fraction of time spent accreting, and (ii) the Eddington ratio, $f_{\rm Edd} = \dot{M}/\dot{M}_{\rm Edd}$.
The time evolution of $M_{\bullet}$ is parameterized as:
\begin{equation}
\frac{dM_{\bullet}}{dt} = {\cal D} f_{\rm Edd} \dot{M}_{\rm Edd} = {\cal D} f_{\rm Edd} \frac{4 \pi G }{c \kappa_T}M_{\bullet} \, .
\label{eq:parametrization}
\end{equation} 
Although the two parameters ${\cal D}$ and $f_{\rm Edd}$ have different physical meanings, it is only their product that is relevant for the mass growth. In Eq. \ref{eq:parametrization}, ${\cal D}$ describes the \textit{continuity} of mass inflow, while $f_{\rm Edd}$ describes the \textit{amount} of mass flowing in.

Transforming the previous equation from time to redshift
%$dM_{\bullet}/dt \times dt/dz = dM_{\bullet}/dz$, 
we obtain the expression:
\begin{equation}
\frac{dM_{\bullet}}{dz} = - \frac{{\cal C} M_{\bullet}}{{\cal E}(z)} \, ,
\end{equation}
where we define:
\begin{equation}
%{\cal C} \equiv 
{\cal C}({\cal D},f_{\rm Edd}) \equiv \frac{4 \pi G {\cal D}f_{\rm Edd}}{H_0 c \kappa_T} \, ,
\end{equation}
and
\begin{equation}
{\cal E}(z) = (1+z)[(1+z)^2(\Omega_0 z+1) - \Omega_{\Lambda}z(z+2)]^{1/2} ~.
\end{equation}
The parameter ${\cal C}$ incorporates constants, the evolution parameters ($f_{\rm Edd}$, ${\cal D}$) and the local value of the Hubble constant $H_0$. 
The values adopted for ${\cal D}$ and $f_{\rm Edd}$ are taken from \cite{PVF_2015, Inayoshi_2016_super, Begelman_2017}; in general, ${\cal D} \sim 1$ and $f_{\rm Edd} \gg 1$ in the high-efficiency region and ${\cal D} \ll 1$ and $f_{\rm Edd} \ll 1$ in the low-efficiency region.

Starting the evolution from redshift $z_0$, corresponding to a seed mass $M_{{\bullet},0}$, we obtain the final expression:
\begin{equation}
M_{\bullet}(z) = M_{{\bullet},0} \exp \left(  \int_{z}^{z_0} \frac{{\cal C} d{\rm z}}{{\cal E}(z)}  \right) ~.
\label{eq:cosmo_evolution}
\end{equation}

We note that the average gas density around the black hole will change with redshift, as a result of local dynamical (galaxy mergers) and thermodynamical (supernovae, outflows) effects.
The exact prediction of the redshift evolution of the mean density of a galaxy requires cosmological simulations. Moreover, the variation is highly dependent on the specifics of the halo history, since each halo will experience a different number of mergers at different cosmic epochs  \citep{Prieto_2017}.
For our calculations we restrict ourselves to the highest redshifts, where it is appropriate to assume that the gas number density ranges between $10^{-2} <n_{\infty}(\mathrm{cm^{-3}})< 10^8$. This range is very large but, as noted by \cite{Inayoshi_2016_super}, this amply takes into account the dynamical diversity of halo histories and the baryonic properties at these cosmic epochs. For the various realizations we draw values randomly from this range to explore statistical outcomes. Our treatment in this work is therefore conservative but idealistic.

The left panel of Fig.~\ref{fig:super-super} shows examples of the cosmological evolution (from $z_0=20$ to $z=7$) of two black hole seeds. In the right panel, for clarity, we zoom-in only on the mass evolution.
The initial mass of the two test seeds is fixed at $M_{\bullet,1} = 10^2 \, \mathrm{\Msun}$ and $M_{\bullet,2} = 10^5 \, \mathrm{\Msun}$, to probe the representative growth tracks for low-mass and high-mass black hole seeds, respectively. The initial density is also set, in order for the low-mass seed to start off in the low-efficiency region ($n_{\infty,1} = 5 \times 10^6 \, \mathrm{cm^{-3}}$) and the high-mass seed in the high-efficiency region ($n_{\infty,2} = 10^7 \, \mathrm{cm^{-3}}$).
This approach is physically motivated. Several works (e.g., \citealt{Pelupessy_2007, Alvarez_2009}) have shown that low-mass seeds are formed in low-density regions, due to the effect of stellar feedback that sweeps the surrounding gas away, causing a delay before the central density is rebuilt. Therefore, low-mass seeds are prone to form in the lower left part of Fig. \ref{fig:super-super}, well outside the high-efficiency region. Instead, the initial density around a high-mass seed is expected to be large (e.g., \citealt{PVF_2015} and references therein).
The seeds are then evolved from $z=20$ to $z=7$, in 100 steps equally spaced in redshift. At each step, the new density is drawn from a uniform distribution in the range $-2<\mathrm{Log}_{10}(n_{\infty}[\mathrm{cm^{-3}}]) < 8$, while the mass is evolved according to Eq. \ref{eq:cosmo_evolution}. This equation depends on the two parameters ($f_{\rm Edd}$, ${\cal D}$). If the black hole seed is inside a high-efficiency (low-efficiency) region, then these parameters are randomly drawn from uniform distributions within the ranges $0.5<{\cal D} \leq 1 $, $1 \leq f_{\rm Edd} < 100$ ($0 \leq {\cal D} \leq 0.5$, $0 \leq f_{\rm Edd} < 1$).

A red line indicates an inefficient growth (with low duty cycles and Eddington ratios), while a green line indicates an efficient growth. A massive seed grows very rapidly to a mass $\sim 2 \times 10^{9} \, \mathrm{\Msun}$ by $z \sim 7$, matching the observation of ULAS J1120+0641 \citep{Mortlock_2011}, see the bottom panel. Several evolutionary runs have also shown seeds reaching masses $> 10^{10} \, \mathrm{\Msun}$ by $z \sim 7$. The accretion is enhanced every time the seed is inside the high-efficiency region, which serves as a ``growth booster". 
A $10^2 \, \mathrm{\Msun}$ seed in the low-efficiency region never experiences fast growth and reaches a final mass of only $\sim 2 \times 10^2 \, \mathrm{\Msun}$ by $z=7$.

We want to make clear that this is not intended as an accurate cosmological simulation of the evolution of black hole seeds. The only purpose of this exercise is to show that the growth of a black hole seed is boosted each time it is situated inside the high-efficiency region.

%%%%%%%%%%%%%%%%%%%%%%%%%%%%%%%%%%%%%%%%%%%%%%%%%%%%%%%%%%%%%%%%%%%%%%
%% SECTION 4:
%%%%%%%%%%%%%%%%%%%%%%%%%%%%%%%%%%%%%%%%%%%%%%%%%%%%%%%%%%%%%%%%%%%%%%
\section{A Probabilistic View of the Growth}
\label{sec:probability}
Our treatment, combining constraints on a range of spatial scales and gas flow properties consistent with conditions available in the early Universe, allows the development of a probabilistic picture for early black hole growth.
The parameter space in Fig.~\ref{fig:super-super} suggests that if a black hole seed forms in the high-efficiency region, it grows very rapidly.
Initial seed masses are not equally likely to occur: they are distributed statistically with an initial mass function (IMF).
Comparing the IMF of black hole seeds with the high-efficiency region in the parameter space $M_{\bullet}-n_{\infty}$, we can further estimate the probability that a seed is able to reach the SMBH stage rapidly.

The IMF of high-$z$ black hole seeds is highly uncertain. Here, we focus on two formation models, DCBH and the Pop III remnants, which lead to massive and light seeds, respectively. DCBHs are predicted to form via the collapse of gas that is free of metals and molecular hydrogen in high-$z$ halos. Pop III stars are the first population of massive and metal-free stars formed.
To model the IMF of DCBHs we fit the results of \cite{Ferrara_2014}, valid for intermediate mass black holes (IMBHs), with a Gaussian distribution:
\begin{equation}
\Phi(\mathrm{DCBH}) = \mathrm{NORM}(\mathrm{Log}\mu = 5.1, \mathrm{Log}\sigma = 0.2) \, ,
\end{equation}
where $\mu$ and $\sigma$ are the mean and the standard deviation, respectively. Here, we are considering the ``fertile model" of \cite{Ferrara_2014}, in the combination super-massive star/DCBH. In our approximation, the bimodality of the Gaussian is not relevant.
For the IMF of Pop III stars, we use a simple model involving a Salpeter-like exponent and a low-mass cutoff $M_{c}$ that depends on the stellar population \citep{Hirano_2014, Pacucci_2017_GW}:
\begin{equation}
\Phi(\mathrm{Pop III},m) \propto m^{-2.35} \exp \left({-\frac{M_{c}}{m}} \right) \, ,
\end{equation}
where $m$ is in solar units. We assume the low-mass cutoff for Pop III stars to be $M_{c} = 10\, \mathrm{\Msun}$ (for Pop I/II stars it would be $\sim 0.35 \, \mathrm{\Msun}$).
We then convolve the IMF  with the relation between the stellar mass and the mass of the remnant \citep{Woosley_2002}, obtaining a gap corresponding to the pair instability supernovae, leaving no remnant.

\begin{figure}
\vspace{-1\baselineskip}
\hspace{-0.5cm}
\begin{center}
\includegraphics[angle=0,width=0.50\textwidth]{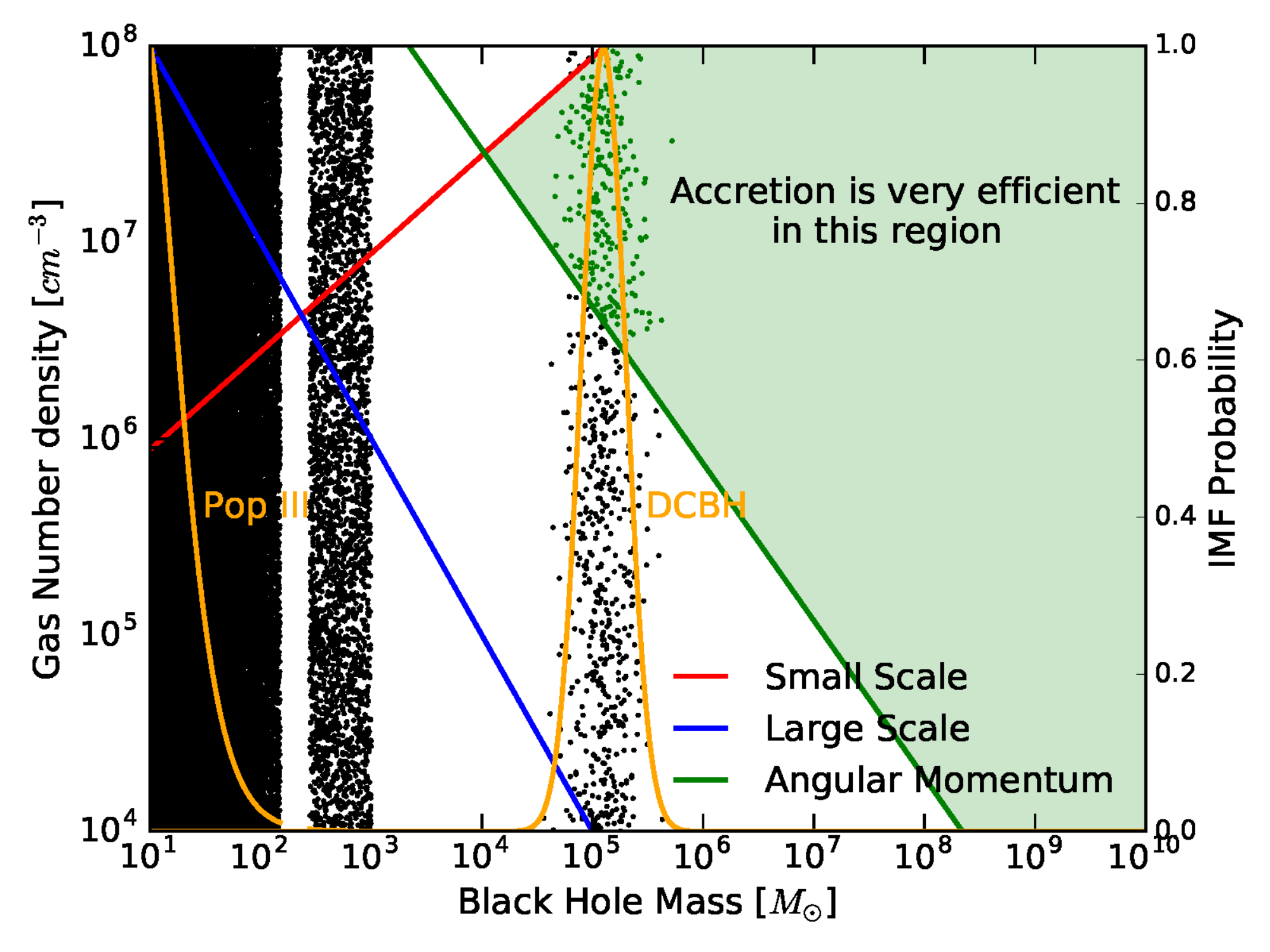}
\caption{Probability that a black hole seed is formed in a high-efficiency region of the $M_{\bullet}-n_{\infty}$ parameter space. The IMFs for Pop III and DCBH seeds are shown with orange lines (right axis labels). In this plot we adopted a value of ${\cal R} = 10^{-3}$.
An ensemble of $10^6$ black hole seeds randomly populates the parameter space (filled points); those falling within the high-efficiency region are green. While the black hole mass follows the IMFs of the two populations of seeds, we assume that the gas number density is equally probable over the range $ 10^4 <n_{\infty}(\mathrm{cm^{-3}})< 10^8$. We estimate the probability that a black hole seed forms directly in the high-efficiency region as a $3\sigma$ upper limit $\lesssim 13 {\cal R}\% = 13 (n_{\rm DCBH}/n_{\rm PopIII})\%$}
\label{fig:super-super_IMF}
\end{center}
\end{figure}

The IMFs for DCBHs and Pop III seeds are shown in Fig.~\ref{fig:super-super_IMF} with orange lines.
While Pop III seeds are too light to be inside the high-efficiency region right at birth, DCBHs are well capable of accreting very efficiently right from the start.
Assuming that Pop III remnants and DCBHs are the representative formation channels for light and massive black hole seeds, we estimate the probability of direct access to the high-efficiency region since birth.

The relative number of Pop III and DCBHs in the high-$z$ Universe is largely unconstrained, but it affects the probability that we compute. To estimate the quantity ${\cal R} \equiv n_{\rm DCBH}/n_{\rm PopIII}$, we employ the luminosity function of high-$z$ quasars (e.g., \citealt{Masters_2012}, which uses the COSMOS photometry). We assume that the high-luminosity end is mainly produced by high-mass seeds, while the low-luminosity end is produced by low-mass seeds, and we extrapolate the results to $z \gsim7$.
The demarcating absolute magnitude $M_{\rm cut}$, above which we assume that the luminosity function is mainly produced by efficient accretors, is also a parameter in our model and it is also observationally unconstrained.
We performed the computation of the probability for different values of the UV magnitude ($\lambda = 1450 \, \mathrm{\AA}$) of the cut $M_{\rm 1450, cut}$ and we express the probability as a function of ${\cal R}$. With the luminosity function of \cite{Masters_2012}, and employing the bolometric corrections in \cite{Runnoe_2012}, we used the following values for the demarcating magnitude: $M_{\rm 1450, cut} = [-25,-26,-27]$. For reference, we note that the break luminosity between the bright and faint ends in \cite{Masters_2012} is  $M^*_{1450} \approx -25.5$.

We generated an ensemble of $10^6$ black hole seeds drawn from the IMFs to randomly populate the $M_{\bullet}-n_{\infty}$ parameter space. We assume that the gas number density is equally probable over the range $ 4 <\mathrm{Log_{10}}(n_{\infty}[\mathrm{cm^{-3}}])< 8$, a realistic assumption for the \textit{formation} of seeds.
We estimate the probability that a black hole seed forms directly in the high-efficiency region as a $3\sigma$ upper limit $\lesssim 13 {\cal R}\% = 13 (n_{\rm DCBH}/n_{\rm PopIII})\%$. Only the DCBH formation channel, or other equivalent methods to form high-mass seeds, comfortably puts a black hole seed into the high-efficiency region since birth. In fact, as the value of ${\cal R} = n_{\rm DCBH}/n_{\rm PopIII}$ approaches zero, the probability becomes negligible as well. High-mass seeds growing efficiently can easily reach SMBH masses of $\gtrsim 10^9 \, \mathrm{\Msun}$ at early times.
To conclude, we note that moderate changes in the shape of both IMFs (Pop III and DCBHs) would have negligible effects on our calculation of the probability.

%

%%%%%%%%%%%%%%%%%%%%%%%%%%%%%%%%%%%%%%%%%%%%%%%%%%%%%%%%%%%%%%%%%%%%%%
%% SECTION 5:
%%%%%%%%%%%%%%%%%%%%%%%%%%%%%%%%%%%%%%%%%%%%%%%%%%%%%%%%%%%%%%%%%%%%%%
\section{Observational Consequences of the Accretion Regimes}
\label{sec:observations}

We now discuss some observational consequences of the accretion regimes described in Fig.~\ref{fig:super-super}, starting from the low efficiency and then moving to the high efficiency.

Black holes born into the low-efficiency region are characterized by low values of the duty cycle ${\cal D}$ and low Eddington ratios $f_{\rm Edd}$. Under these conditions, accreting black holes will be unobservable, since: (i) their bolometric luminosities $L$ will be low, being proportional to their Eddington ratios, and (ii) they would be accreting only for a small fraction of time, since their duty cycles ${\cal D}$ are also low. An inefficient accretor of $10^4 \, \mathrm{\Msun}$ (the highest mass for this class of seeds, independent of the gas density), embedded in an environment with an average gas number density ($10^4 \, \mathrm{cm^{-3}}$), is predicted to shine with a luminosity $\lsim 10^{41} \, \mathrm{erg \, s^{-1}}$, more than $10^6$ times fainter than ULASJ1120+0641 \citep{Mortlock_2011}. 
It has been claimed that we are currently missing faint quasars \citep{Comastri_2015, Pezzulli_2017} in deep surveys due to large obscuring column densities in their host galaxies. Our calculation suggests that, in addition to obscured quasars, there ought to exist another population of undetectable black holes due to their being inefficient accretors, with luminosities $\lsim 10^{41} \, \mathrm{erg \, s^{-1}}$. Evidence for the existence of this population is likely to come from gravitational wave detections of their subsequent merging activity \citep{Sesana_2007, Tanaka_2009}.

Conversely, the detection of a black hole accreting at super-Eddington rates in the high-$z$ Universe, we argue, would convey more information about its environmental conditions, namely $M_{\bullet}$ and $n_{\infty}$, as it is predicted to lie in the high-efficiency locus of Fig.~\ref{fig:super-super}. The occurrence of super-Eddington accretion in the early Universe could also be established by the presence of relativistic jets, as several studies have suggested (e.g., \citealt{Ghisellini_2013}).
In addition, in the view of a population study, we predict the high-luminosity end of high-$z$ quasar luminosity functions ($z \gsim 10$) to be populated preferentially by black holes in the high-efficiency region, which are easier to detect. Since high-mass black hole seeds are a high-redshift phenomenon only, we could expect the $z \gsim 10$ quasar luminosity function to be significantly different from the one at lower redshifts.
%%%%%%%%%%%%%%%%%%%%%%%%%%%%%%%%%%%%%%%%%%%%%%%%%%%%%%%%%%%%
%% SECTION 6: Discussion and Conclusions
%%%%%%%%%%%%%%%%%%%%%%%%%%%%%%%%%%%%%%%%%%%%%%%%%%%%%%%%%%%%
\section{Discussion and Conclusions}
\label{sec:disc_concl}
Sustaining large accretion rates for an extended period of time is a fundamental requirement to form SMBHs early in the evolution of the Universe, in order to match the observations of $\sim 10^9 \, \mathrm{\Msun}$ objects at $z \sim 7$.
In this Letter, we investigate the environmental conditions of the system (black hole + host galaxy) that lead to optimal growth of black holes at early epochs, in the context of theoretically expected cosmological conditions permitted within current structure formation models.
By combining conditions on small scales ($r \ll R_B$) and large scales ($r \gsim R_B$), taking into account the angular momentum of the host galaxy, we describe the optimal conditions in the $M_{\bullet}-n_{\infty}$ parameter space that lead to a continuous stream of gas flowing unimpeded from large scales down to the black hole. This would likely lead to the formation of the highest SMBH masses in the shortest time. 
%\textbf{The large and small spatial scales are not independent in our model, they are rather connected via a global flow model, based on the assumption of an isothermal density profile for the host halo gas and angular momentum properties of the gas (see Eq. \ref{eq:density}).}

We identify a region with mass $\gsim 10^4 \, \mathrm{\Msun}$ in which accretion is very efficient and proceeds continuously (${\cal D} \sim 1$) with large accretion rates ($f_{\rm Edd} \gg 1$). Under these conditions, the growth of black holes to $\gsim 10^9 \, \mathrm{\Msun}$ to account for the brightest detected quasars at high redshift is highly feasible. On the contrary, for low-mass black holes ($\lesssim 10^4 \, \mathrm{\Msun}$) accretion is very inefficient.
By assuming (i) a distribution of masses for both DCBHs and Pop III black hole seeds, (ii) a relative fraction ${\cal R} = (n_{\rm DCBH}/n_{\rm PopIII})$ of low-mass and high-mass seeds, and (iii) a uniform distribution in gas density at the formation of the seed $4 <\mathrm{Log_{10}}(n_{\infty}[\mathrm{cm^{-3}}])< 8$, we find that the probability of black hole seeds being born directly inside the high-efficiency region has a $3\sigma$ upper limit of $ 13 {\cal R}\% = 13 (n_{\rm DCBH}/n_{\rm PopIII})\%$. The uncertainty in the probability estimate is related to the particular threshold value above which we assume that the high-$z$ luminosity function is determined by efficient accretors. 
Our analysis suggests that black hole seeds with high initial masses begin accreting at high efficiency. Therefore black hole seeds can be rapid growers if they form in a high-efficiency region, or be extremely inefficient growers otherwise. 
This picture suggests that a large population of high-$z$ lower-mass black holes that formed in the low-efficiency region might remain undetectable as quasars, since we predict their bolometric luminosities to be $\lesssim 10^{41} \, \mathrm{erg \, s^{-1}}$. In the presence of mergers between black holes, seeds can of course be efficient growers even in the low-efficiency region.

Our study does not pinpoint or require any particular formation channel for high-$z$ black holes but is dependent only on the initial mass of the seed.
It is possible that low-mass seeds start out as inefficient growers but enter the high-efficiency region at a later time.
For instance, dynamical mechanisms (see, e.g., \citealt{Alexander_2014}) can rapidly boost an initial seed of $\sim 10^2 \, \mathrm{\Msun}$ to about $\sim 10^4 \, \mathrm{\Msun}$, putting it very close to the high-efficiency region at early cosmic times. 
Of course, DCBH seeds that start out already in the high-efficiency region are the most likely progenitors of $z \sim 7$ SMBHs.
For the same reason, however, it is not straightforward to discern observationally light and massive seeds by $z \sim 7$, because by that epoch the initial conditions of the distribution of black holes have been erased. Inefficient growers could transit into high-efficiency regions and grow very rapidly thereafter.
From Fig.~\ref{fig:super-super}, however, one solid conclusion can be drawn: \textit{once a seed is in a high-efficiency region, its growth will likely be gas supply-limited}. In fact, if the gas density does not change dramatically over short times, the black hole mass is bound to increase, and this pushes it further inside the high-efficiency region. 
Therefore, for such seeds the feedback from the black hole itself is not going to be a showstopper. Only significant changes to the density arising from external factors (e.g., stellar feedback, galaxy merger, passage of a gas cloud) can radically perturb its growth trajectory.

JWST detections of low-luminosity high-$z$ quasars will be crucial to understand whether optimal conditions for black hole growth are rare or ubiquitous in the early Universe. These observations will have important implications for our understanding of early galaxy formation as well.

\vspace{0.3cm}
F.P. acknowledges useful discussions with Kohei Inayoshi and Andrea Ferrara.
F.P. and N.C. acknowledge the Chandra grant No. AR6-17017B.
N.C. acknowledges NASA-ADAP grant No. NNX16AF29G, Chandra grant No. GO5-16150A and NASA/12-EUCLID11-0003.
P.N. acknowledges the TCAN grant No. 1332858.

%%%%%%%%%%%%%%%%%%%%%%%%%%%%%%%%%%%%%%%%%%%%%%%%%%%%%%%%%%%%%%%%
%%  REFERENCES
%%%%%%%%%%%%%%%%%%%%%%%%%%%%%%%%%%%%%%%%%%%%%%%%%%%%%%%%%%%%%%%%

\end{document}